%% file: ms.tex
\documentclass[usenatbib,useAMS,a4paper,usenames,dvipsnames]{mn2e}
\usepackage{color}
\usepackage[breaklinks]{hyperref}
\definecolor{darkblue}{rgb}{0.0,0.0,0.3}
\hypersetup{colorlinks,breaklinks,
            linkcolor=darkblue,urlcolor=darkblue,
            anchorcolor=darkblue,citecolor=darkblue}
\usepackage{txfonts}

\DeclareSymbolFont{cmletters}{OML}{cmm}{m}{it}
\DeclareMathSymbol{v}{\mathalpha}{cmletters}{"76}
\usepackage{graphicx}

\newcommand{\useiop}{-3}

\newcommand{\shortauthors}[1]{}
\newcommand{\shorttitle}[1]{}
\newcommand{\altaffiltext}[2]{}
\newcommand{\eqref}[1]{(\ref{#1})}
\input{macros}
\usepackage{ifthen}
\definecolor{MyDarkBlue}{rgb}{0,0.08,0.7}

\defcitealias{tch08}{TMN08}
\defcitealias{tch09}{TMN09}
\defcitealias{tch11}{TNM11}
\defcitealias{kom09}{K09}
\defcitealias{bz77}{BZ}
\defcitealias{bp82}{BP}

\newcommand{\OmegaH}{\Omega_\mathrm{H}}

\newcommand{\rH}{r_\mathrm{H}}
\newcommand{\Pran}{\mathrm{Pr_m}}

\newcommand{\const}{{\rm constant}}

\newcommand{\gdet}{\sqrt{-g}}

\newcommand{\avg}[1]{\ensuremath{\langle#1\rangle}} 
\newcommand{\abs}[1]{\ensuremath{\left|#1\right|}}

\newcommand{\MdotH}{\dot M}

\newcommand{\etaj}{\eta_{\rm jet}}
\newcommand{\etaw}{\eta_{\rm wind}}

\newcommand{\etabz}{\eta_{\rm BZ}}

\newcommand{\phibh}{\phi_{\rm BH}}
\newcommand{\PhiBH}{\Phi_{\rm BH}}

\newcommand{\cut}[1]{\hbox{}}

\shortauthors{}
\shorttitle{}

\ifthenelse{\equal{\useiop}{-3}}{ 

\author[A.~Tchekhovskoy 
and
J.~C.~McKinney]
{Alexander Tchekhovskoy$^1$\thanks{\hbox{E-mail:
      atchekho@princeton.edu~(AT)}} and
Jonathan C. McKinney$^2$
\\
  $^1$Center for Theoretical Science, Jadwin Hall, Princeton University, Princeton,
  NJ 08544; Princeton Center for Theoretical Science Fellow \\
 $^2$Kavli Institute for Particle Astrophysics and Cosmology, Stanford University, P.O. Box 20450, MS 29,
Stanford, CA 94309}
}{
}
\begin{document}
\label{firstpage}

\title[Jets from Prograde and Retrograde Black Holes]%
{\hbox{Prograde and Retrograde Black Holes: Whose Jet is More Powerful?}}

\ifthenelse{\equal{\useiop}{-2}}{
\journal{New Astronomy}
\begin{frontmatter}

\author[cfa]{Alexander Tchekhovskoy}
\ead{atchekho@princeton.edu}

\address[cfa]{Princeton Center for Theoretical Science, Jadwin Hall, Princeton
  University, Princeton, NJ 08544, USA}
}{}
\ifthenelse{\equal{\useiop}{3}}{
\author{Alexander Tchekhovskoy$^1$ and $^2$} 
  \maketitle
  \begin{affiliations}
    \item Princeton Center for Theoretical Science, Princeton
  University, Jadwin Hall, Princeton
  NJ 08544; atchekho@princeton.edu\\
  \end{affiliations}
}{}
\ifthenelse{\equal{\useiop}{0}}{
\altaffiltext{1}{Princeton Center for Theoretical Science, Princeton
  University, Jadwin Hall, Princeton
  NJ 08544; atchekho@princeton.edu}
}

\ifthenelse{\equal{\useiop}{-3}}{ 
\date{Accepted . Received ; in original form }
\pagerange{\pageref{firstpage}--\pageref{lastpage}} \pubyear{2012}
\maketitle
}

\begin{abstract}

  The outflow efficiency ($\eta$) from black hole (BH) accretion disc
  systems is known to depend upon both the BH spin ($a$) and the
  amount of large-scale magnetic flux threading the BH and disc.
  Semi-analytical flux-trapping models suggest retrograde BHs should
  trap much more large-scale magnetic flux near the BH leading to much higher
  $\eta$ than for prograde BHs.  We self-consistently determine the
  amount of large-scale magnetic flux trapped by  rapidly spinning
  ($a=-0.9$ and $0.9$) BHs using global 3D time-dependent
  non-radiative general relativistic magnetohydrodynamic simulations
  of thick ($h/r\approx0.3{-}0.6$) discs. 
  We find that BH-trapped flux  builds up until
  it is strong enough to disrupt the inner accretion disc.
  Contrary to prior flux-trapping models, which 
  do not include 
  the back-reaction of magnetic flux on the disc, our simulations show
  prograde BHs trap more magnetic flux, leading to about
  $3$ times higher~$\eta$ than retrograde BHs for $\abs{a}=0.9$.  Both spin
  orientations can produce highly efficient jets, $\eta\sim100$\%,
  with increasing $\eta$ for increasing disc thickness.  The
  similarity of $\eta$ for prograde and retrograde BHs makes it
  challenging to infer the sign of $a$ based on jet energetics alone.

\end{abstract}

\ifthenelse{\equal{\useiop}{-2}}{
\end{frontmatter}
}{}

\ifthenelse{\equal{\useiop}{-3}}{ 
}{}
\ifthenelse{\equal{\useiop}{3}}{
  }{}
  
\ifthenelse{\equal{\useiop}{0}}{
{
  }
}


\section{Introduction}
\label{sec:introduction}

\begin{figure*}
  \begin{center}
    \includegraphics[width=0.87\textwidth]{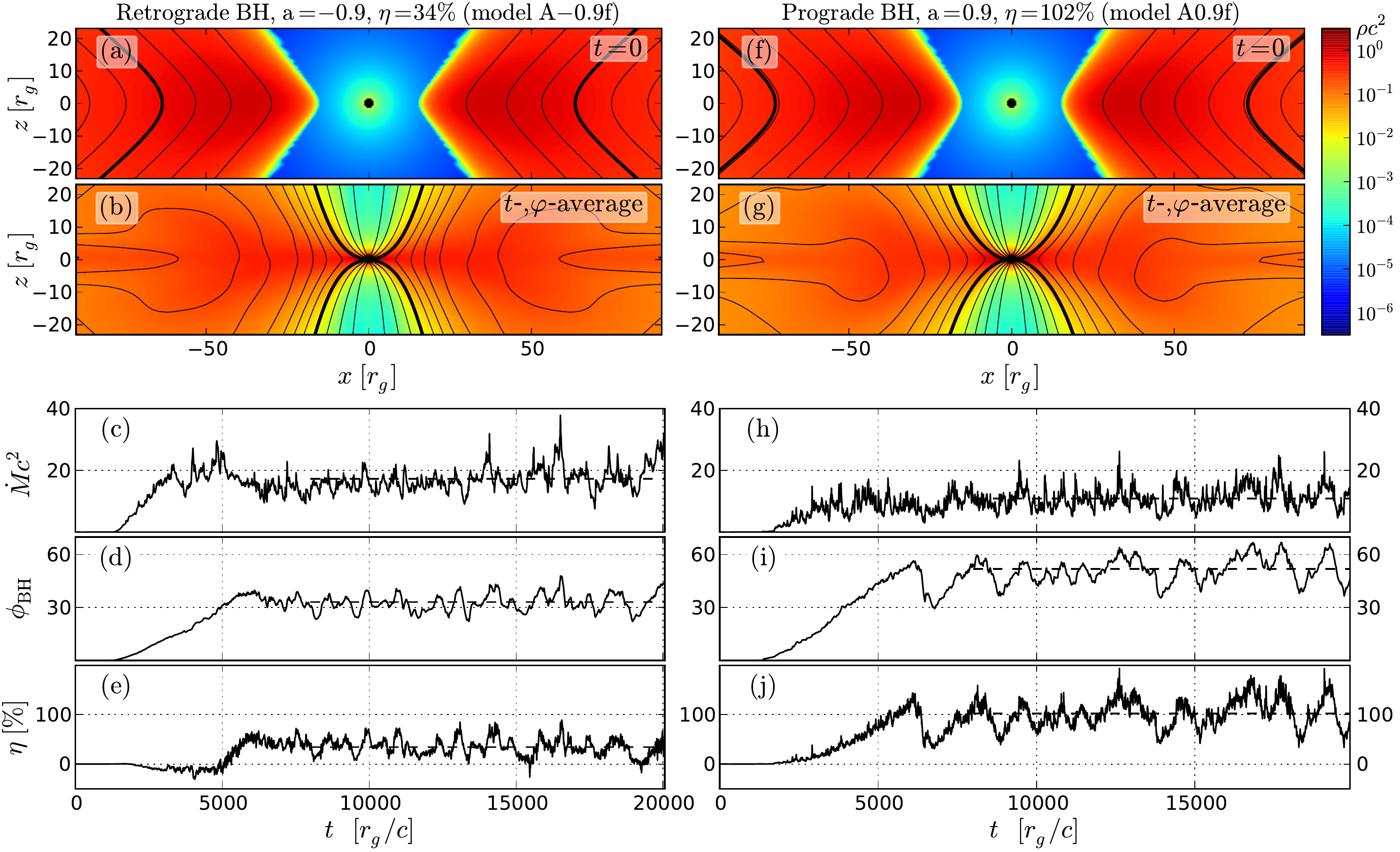}
  \end{center}
  \vspace{-0.2cm}
  \caption{Initial conditions and time-evolution of our fiducial models, 
    $\mathrm{A{-}0.9f}$ (left column) and $\mathrm{A0.9f}$
    (right column). See Supporting Information for movies.
    [Panels a,f] A vertical slice through the initial
conditions. Colour shows fluid frame rest-mass energy density (see
colour bar) and thin black lines show 
levels of constancy of enclosed magnetic flux, $\avg\Phi$, which represent field lines in the
image plane.  Magnetic field is axisymmetric, with
$B_\varphi=0$ everywhere and $B_{\rm
  z}>0$ out to $\textrm{few}\times100r_g$.  
Only the magnetic flux enclosed by the thick solid field line
eventually falls into the BH. [Panels b,g]
Show $t$- and $\varphi$-average of the
magnetically-arrested state of the simulation. Solid lines show the same
contours of $\avg\Phi$ as in panels (a) and (f). Accretion accumulates so much 
flux in the centre that the inner disc is able to push 
only a fraction of the flux, enclosed by the thick
line, into the BH. 
[Panels c-e,h-j] From top to bottom: rest-mass
energy accretion rate, $\dot M c^2$, dimensionless BH magnetic
flux, $\phibh$, and energy outflow efficiency, $\eta$. 
Both $\phibh$ and $\eta$ saturate ar $t
\gtrsim 6000 r_g/c$, beyond which
the accretion flow is magnetically arrested.
Dashed lines show time-averages: a prograde BH has a larger efficiency, $\eta=102$\%, than
a retrograde BH, $\eta=34$\%.
}
\label{fig:mkmdot}
\end{figure*}

Various astrophysical systems---active
galactic nuclei (AGN), bla\-ck hole binaries (BHBs), and gamma-ray
bursts (GRBs)---produce relativistic outflows. Mediated by large scale
magnetic fields, outflows can be powered by the central BHs via 
\citet[\citetalias{bz77} hereafter]{bz77} mechanism
or by the inner regions of 
accretion discs via \citet[\citetalias{bp82} hereafter]{bp82} mechanism. 

Large-scale
magnetic fields
extract BH spin energy 
at a rate,
\begin{equation}
  \label{eq:bz6}
  P_{\rm BZ} = \frac{\kappa}{4\pi c} \PhiBH^2 \OmegaH^2\, f(\OmegaH),
\end{equation}
where $\kappa\approx0.05$
weakly depends on 
field geometry,
$\PhiBH
= (1/2)\iint_{\theta,\varphi} \abs{B^r}\mathrm{d}A_{\theta\varphi}$ is
an absolute magnetic flux through the BH, with
the integral over all $\theta$ and $\varphi$ on the
BH horizon, $r_{\rm H}=r_g(1+\sqrt{1-a^2})$. Here
$r_g=GM/c^2$, $M$, and $a$  are BH gravitational radius,
mass, and dimensionless spin,
$\OmegaH=ac/2\rH$ is BH angular frequency,
$f(\OmegaH) \approx 1+1.38(\OmegaH r_g/c)^2-9.2(\OmegaH r_g/c)^4$ is a
high-spin correction, 
$\mathrm{d} A_{\theta\varphi}=\gdet \, \mathrm{d}\theta
\mathrm{d}\varphi$ is an area element in $\theta-\varphi$ plane, and
$g$ is the metric determinant 
\citep[see also \citetalias{bz77}; \citealt{lwb00}; \citealt{kom01}; \citealt{tn08}]{tch10a}.

For fixed $\PhiBH$, $P_{\rm BZ}$ is the same for 
prograde ($a>0$) and retrograde ($a<0$) BHs.  However, 
a massive accretion disc introduces a preferred spin/rotation direction.
Does this lead to large differences in $\PhiBH$ and jet powers of prograde and
retrograde BHs? Understanding this is crucial, as the link between 
jet power and BH spin has been confirmed in BHBs 
\citep[but see \citealt{fender2010}]{nm12} and
jet power is increasingly often used for inferring BH spin
\citep{daly2011,mr11a,gne11,lz11,bam12}.

To quantify jet strength in an accretion system, we define BZ efficiency 
as jet power (eq.~\ref{eq:bz6}) 
in units of mass accretion rate, $\MdotH$,
\begin{equation}
  \label{eq:etabz}
  \etabz = \frac{{P_{\rm BZ}}}{\avg{\MdotH}c^2} \times 100\%
         = \frac{\kappa}{4\pi} {\phibh^2}\left(\frac{\OmegaH
             r_g}{c}\right)^2\, f(\OmegaH) \times 100\%,
\end{equation}
where $\phibh= \PhiBH/(\avg{\MdotH} r_g^2 c)^{1/2}$ is BH
dimensionless magnetic flux, and $\avg{...}$ is a time average.
Similarly, we define magnetic flux enclosed by a toroidal ring,
$(r,\theta)$, via $\Phi(r,\theta)=\iint_{\theta'<\theta,\varphi'}B^r
dA_{\theta'\varphi'}$, and its
dimensionless version, $\phi(r,\theta)=\Phi(r,\theta)/(\avg{\MdotH}
r_g^2 c)^{1/2}$, where the integral is over
a polar cap, $\theta'<\theta$, of a sphere of radius $r$.

Clearly, jet efficiency depends on the ability of accretion flows to
drag large-scale magnetic fields to the centre. 
\citet{lubow1994} studied magnetic field dragging by geometrically thin accretion
discs, with angular thickness, $h/r\ll1$. 
They found that field transport is governed by the
value of the magnetic Prandtl number ($\Pran$, which is the ratio of turbulent
viscosity to resistivity) and that in 
order for field dragging to be efficient, this number must be
large, $\Pran\gtrsim (h/r)^{-1}\gg1$.  
Magneto-rotational instability (MRI, \citealt{bal91}) driven
turbulent accretion flows have low
values of magnetic Prandtl number, $\Pran\sim1$
\citep{fs09,gg09,ll09}. For such natural values, $\Pran\sim1$,
thin discs do not 
produce centrally-concentrated magnetic fields, which suggests that thin discs 
produce weak or no jets  (\citealt{lubow1994}; but see \citealt{spruit_uzdensky_2005,rl08}).
To include
general-relativistic 
(GR) effects of the innermost stable circular orbit (ISCO)
on field dragging in thin discs, 
\citet{rgb06}
proposed that gas and fields plunge into the BH inside the ISCO, so
no magnetic flux passes through the ``gap'' between the
BH horizon and the ISCO. Based on a GR-version of this flux-trapping ``gap'' model,
formally applicable only to thin discs, \citet{gar09}; \citet{ges10} 
concluded that 
jets from retrograde BHs, $a=-0.9$, are $\gtrsim10\times$ more powerful than from
prograde BHs, $a=0.9$, both for thin and thick discs
(see \S\ref{sec:results}).

\begin{table*}
\begin{center}
\caption{Simulation details}
\begin{minipage}{\textwidth}
\begin{center}
  \begin{tabular}{l@{$\quad$}c@{$\quad$}c@{$\quad$}c@{$\quad$}c@{$\quad$}c@{$\quad$}c@{$\quad$}c@{$\quad$}c@{$\quad$}c@{$\quad$}c@{$\quad$}c@{$\quad$}c}
\hline
Name$^a$ &  $a$  & $\eta^b\ [\%]$ & $\beta_{\rm min}$ & $\Delta\varphi$ &
Resolution$^c$ & $r_{\rm in}/r_g$ &
$r_{\rm max}/r_g$ &$R_{\rm in}/r_{\rm H}$& $R_{\rm out}/r_g$ 
& $r_{\rm br}/r_g$ & $t_{\rm run}\ [r_g/c]$ & $t_{\rm avg}\ [r_g/c]$\\
\hline
\multicolumn{13}{|c|}{Simulations with $(h/r)_{r_{\scriptstyle\rm max}}\approx0.2$
  and $\avg{h/r}_{30}\approx0.3$:}\\
\input{simtex.txt}
\hline
\multicolumn{13}{|c|}{Simulations with $(h/r)_{r_{\scriptstyle\rm max}}\approx0.6$
  and $\avg{h/r}_{30}\approx0.6$:}\\
\input{simtex_thick.txt}
\hline
\hline
\label{tab1}
\end{tabular}
\end{center}
\vspace{-0.5cm}
$^a$ Suffixes ``$h_\theta$''
or ``$l_\theta$'' indicate that $\theta$-resolution 
was increased or, respectively, decreased by a factor of two as compared to the
fiducial model. Models A0.9$l_\theta$, A0.9$h_\theta$, and
A0.9$h_\theta h_\varphi$ are similar to model A0.9 but with $\theta$- and
$\varphi$-resolutions increased by factors $f_\theta$ and
$f_\varphi$, respectively,  at $t=14207r_g/c$, where the pairs of
values, $(f_\theta,f_\varphi)$, are given by  
$(0.5,1)$, $(2,1)$, and $(2,2)$, respectively. 
Model A$-0.9$flip is similar to
model A0.9, but with BH spin value reversed to $a=-0.9$ at $t=14207r_g/c$.
Model A$-0.9h_\theta$ (A$-0.9h_\varphi$) 
is similar to model A$-0.9$ but with 
$\theta$-resolution ($\varphi$-resolution) increased by a factor of two
at $t=12328r_g/c$ (at $t=8000r_g/c$, respectively). Model
A0.9$h^2_\varphi$ is similar to model $A0.9h_\varphi$ but with
$\varphi$-resolution increased by a factor of two at $t=8500r_g/c$.
\\
$^b$ We quote $95.4$\% (two sigma) confidence intervals.  
\ \ $^c$ Given as $N_r\times N_\theta\times N_\varphi$, where $N_r$, $N_\theta$,
and $N_\varphi$ are grid resolutions in $r$-, $\theta$-, and
$\varphi$-directions, respectively.
\end{minipage}
\end{center}
\end{table*}

Geometrically
thick
discs can efficiently drag
large-scale fields inward even for $\Pran\sim1$ \citep{cao11}, and
time-dependent simulations of \citet{mck04,mck05,hk06} show that
thick accretion discs
($h/r\simeq0.2{-}0.3$) can 
produce powerful jets. However, the simulated
values of jet efficiency have a high degree of scatter, e.g.,
jet efficiency from retrograde BHs
ranges  in these works  from $10$\% to $50$\% of the
corresponding efficiency for prograde BHs at the same absolute value
of spin. 
A major uncertainty in such studies is the
dependence of jet efficiency on the value of large-scale vertical magnetic flux
initially present in the disc, which is a free
modeling parameter with no obvious natural value.
Changes in the flux can significantly affect
simulated jet efficiencies 
\citep{mck04,mck05,bhk07} and render them unreliable.

Is it at all
possible to obtain a well-defined value of jet efficiency, 
free from the uncertainties in large-scale 
magnetic flux content of the initial simulation setup? \citet[\citetalias{tch11} hereafter]{tch11} showed that a 
promising approach is to start with
a large vertical magnetic flux in the disc, more than the accreting gas can push 
into the BH. The 
excess flux remains outside,
impedes the accretion, and leads to
a magnetically-arrested disc (MAD,
\citealt{bkr74,nia03,igu03,igu08}; \citetalias{tch11}). 
Estimates show that many astrophysical systems contain enough large-scale magnetic
flux to naturally form MADs \citep{nia03,mtb12}.  
The inner disc properties of MADs
are shown to be \emph{independent of the initial value of large-scale magnetic
  flux}.
So, we can reliably determine $\eta$
for prograde and retrograde BHs and thereby test the flux-trapping
``gap'' model.
 In \S\ref{sec:results} we
describe our numerical method and present our results 
and~in~\S\ref{sec:conclusions}~we~conclude.

\section{Numerical Method and Results}
\label{sec:results}

We carried out time-dependent 3D general relativistic
non-radiative MHD simulations using a high-order version of
Godunov-type shock-capturing code HARM
\citep{gam03,mck04,tch07,tch09,mb09} in
modified spherical polar coordinates.   We use
logarithmically spaced radial grid, $dr/r=\const$, for $r\lesssim
r_{\rm br}$ (see Table~\ref{tab1} for $r_{\rm
  br}$ values). For $r\gtrsim r_{\rm br}$, the grid becomes
progressively sparser, $dr/r=4(\log r)^{3/4}$, with a smooth
transition at $r_{\rm br}$. We choose grid inner radius, 
$R_{\rm in}$, such that there are at least $9$ grid cells between the inner
radial boundary and the BH horizon and place the outer radial boundary at
$R_{\rm out}=10^5 r_g$, which is larger than the
light travel distance in a duration of the simulation (see
Table~\ref{tab1} for $R_{\rm in}$ values).  This ensures that
both radial boundaries are causally disconnected.  
We use standard boundary
conditions (see \citetalias{tch11}): outflow in $r$-,
reflecting in $\theta$-, and periodic in $\varphi$-directions.

We consider a retrograde fiducial model, A$-0.9$f, for spin $a =
-0.9$, and a prograde fiducial model, A$0.9$f, for spin $a=0.9$. 
Figure~\ref{fig:mkmdot}(a),(f) shows our initial
conditions: BH at the centre of an equilibrium hydrodynamic torus
\citep{chakrabarti1985,dev03a}, with
angular velocity, $\Omega\propto r^{-1.75}$. We place the torus inner edge
at $r_{\rm in} = 15 r_g$ and 
pressure maximum at $r_{\rm max}\sim35r_g$, so
disc angular thickness at $r_{\rm max}$ is
$(h/r)_{r_{\scriptstyle\rm max}}\approx0.2$   (see
Table~\ref{tab1}). Here 
$(h/r)_{r}=\bigl[\iint_{\theta,\varphi} (\theta-\pi/2)^2 \rho\,
\mathrm{d}A_{\theta\varphi}\bigr/\!\iint_{\theta,\varphi}\rho\,
\mathrm{d}A_{\theta\varphi}\bigr]^{1/2}$ and the
integral is over all $\theta,\varphi$ on a sphere of radius
$r$. 
We embed the torus with a purely poloidal ($B_\varphi=0$) loop of a weak magnetic field, with plasma $\beta\equiv p_{\rm gas}/p_{\rm
  mag}\ge\beta_{\rm min}=100$. 
The loop contains a large magnetic flux due to the large radial extent of
the loop (see \citetalias{tch11}).

We define rates of accretion of rest mass, $\dot{M}$, and rest mass energy,
$F_M\equiv \dot{M}c^2$,  by $\dot M = -
\iint_{\theta,\varphi} \rho u^r dA_{\theta\varphi} \equiv F_{\rm
  M}/c^2>0$, where $\rho$ is fluid-frame mass density, $u^r$ is
$r$-component of contravariant $4$-velocity,
and the integration is over all $\theta,\varphi$ on the BH horizon,
$r=r_{\rm H}$.  
Figure~\ref{fig:mkmdot}(c),(h) shows  that after the start of the simulation
$\MdotH$ 
settles to a steady state at $t\gtrsim5000r_g/c$.
BH magnetic flux, $\phibh$, increases
until $t\sim6000r_g/c$, beyond which $\phibh$ saturates 
(Figure~\ref{fig:mkmdot}d,i). 
Flux accumulates outside the BH, impedes the accretion, and
leads to a magnetically-arrested accretion disc (MAD,
\citealt{nia03}). 
Some of the BH flux occasionally escapes from the BH via magnetic interchange
(\citealt{ss01}; \citetalias{tch11}), leading to
oscillations of $\phibh$ in time (Fig.~\ref{fig:mkmdot}d,i).
Figure~\ref{fig:mkmdot}(b),(g) shows the $t$- and $\varphi$-average of the flow
over the MAD period ($t_{\rm avg}$ in Table~\ref{tab1}). In order to get to the
BH, gas ``diffuses''
through the vertical magnetic flux that
reached size $\sim25r_g$ in a duration of our simulations. 
At large 
$r\gtrsim30r_g$, all our discs 
have very similar values of angular thickness, $\avg{h/r}_{30}\approx0.3$ 
(see Table~\ref{tab1}). At smaller $r$, funnel magnetic fields compress the
disc vertically and lead to smaller $h/r$.

We define total energy accretion rate (as
measured at infinity), $F_E = 
\iint_{\theta,\varphi} {T^r}\!_t \, dA_{\theta\varphi}$, where ${T^\mu}\!_\nu$ is
stress-energy tensor, 
the integral
is over all $\theta,\varphi$ on BH horizon, and
$F_E$ is \emph{positive} if energy flows \emph{into} the BH. 
We define energy outflow efficiency $\eta$ as the
energy return rate to infinity
divided by the time-average mass accretion rate: 
\begin{equation}
\eta \equiv \frac{F_M-F_E}{\avg{F_M}} \times 100\%.
\label{eq:eta2}
\end{equation}
Time-dependence of $\eta$ for our fiducial models, A$-0.9$f and
A$0.9$f, is shown in
Figure~\ref{fig:mkmdot}(e),(j). The outflow efficiency
saturates at $t\gtrsim6000r_g/c$ and clearly
correlates with $\phibh$. This is not surprising since we will see
below that most of the outflowing
energy is carried by BZ-driven jets, hence
$\eta\approx\etabz\propto\phibh^2$ (eq.~\ref{eq:etabz}).
A retrograde BH traps $\simeq30$ units of $\phibh$ (panel d), while a prograde BH traps $\simeq50$
units (panel i). Correspondingly, the retrograde BH produces outflows
$(50/30)^2\approx3$ times less efficiently (see eq.~\ref{eq:etabz}), 
with $\eta=34\pm3$\% (panel~e), than the
prograde BH, 
which has $\eta=102\pm10$\% (panel~j and Table~\ref{tab1}). 
It is interesting to compare these
results to those for thicker discs, $\avg{h/r}_{30}\approx0.6$, and a somewhat
larger magnitude of BH spin, $\abs{a}=0.9375$. 
The retrograde model has $\eta=88\pm75$\% 
and the prograde model has
$\eta=354\pm43$\% (see Table~\ref{tab1} and \citealt{mtb12}). 
Thicker discs 
reach higher $\eta$,
and prograde BHs have higher $\eta$ than retrograde BHs.

We repeated our fiducial models
in a reduced azimuthal wedge,
$\Delta\varphi=\pi$. We refer to them as A$-0.9$ and A$0.9$ and
find $\eta=43\pm8$\% and $\eta=96\pm17$\%, respectively,
in agreement with our fiducial models. 
We carried out a number of resolution
studies 
(see Table~\ref{tab1}) and find they agree with our fiducial models
within $\eta$ measurement uncertainty (see Table~\ref{tab1}):
the mean efficiency of our retrograde models is
$\eta=38\pm4$\% and of
prograde models is $\eta=106\pm7$\%. 
Hence, $\eta$ values are converged in our fiducial models, which
at $r=10r_g$ have $Q_z\simeq95$ and
$Q_\varphi\simeq20$ cells per the fastest growing
$z$- and $\varphi$- MRI wavelengths, respectively, and
an equatorial cell aspect ratio, $\delta
r\!:\!r\delta\theta\!:\!r\delta\varphi\approx2\!:\!1\!:\!7$.
Our highest $\varphi$-resolution model, A$0.9h^2_\varphi$, has
$Q_z\approx Q_\varphi\approx100$ and $\delta
r\!:\!r\delta\theta\!:\!r\delta\varphi\approx2\!:\!1\!:\!2$ 
and is
very well resolved according to MRI resolution criteria
 \citep[e.g.,][]{hgk11,srsb12}.

\begin{figure}
  \begin{center}
    \includegraphics[width=0.95\columnwidth]{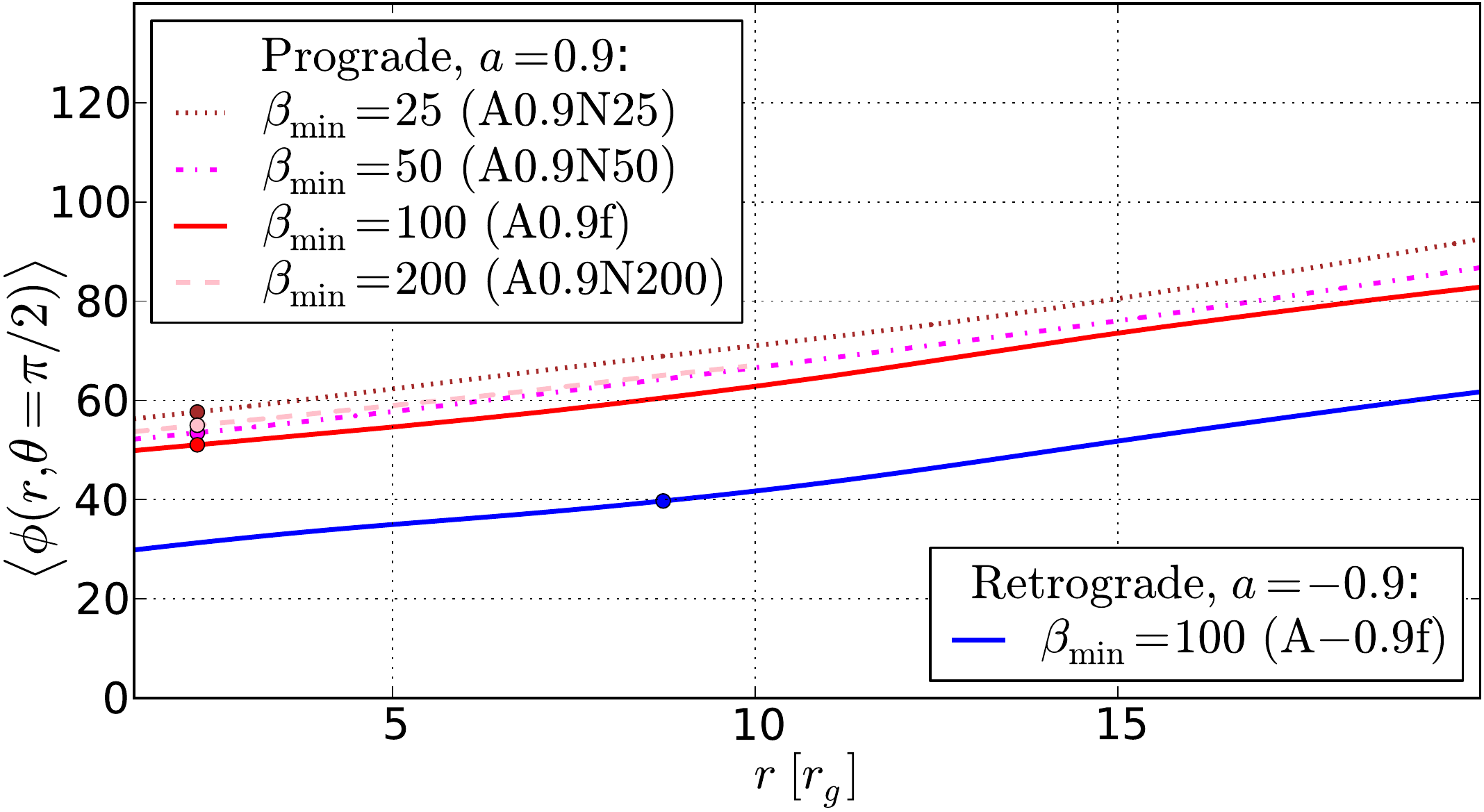}
  \end{center}
  \vspace{-0.2cm}
  \caption{Equatorial profile of magnetic flux, $\avg{\phi(r,\theta=\pi/2)}$, in
    models different by the initial amount of large-scale
    magnetic flux (see legend).  The ISCO radius is indicated
    by filled circles.  Despite the magnetic flux content of the models
    varying by a factor of $\approx3$, all prograde models (top cluster
    of curves) agree to $\lesssim10$\%. Hence, flux accumulation
    around the centre in our simulations
    is independent of the initial large-scale
    flux content of the disc.}
\label{fig:flux}
\end{figure}

Why do outflow efficiencies of prograde and
retrograde BHs differ? Can this be due to differences in initial
conditions amplified by accretion flow turbulence?  
To test this, 
we ran the prograde model A$0.9$
until $t=14207r_g/c$ and instantaneously flipped the direction of BH
spin. We refer to this model as A$-0.9$flip, which gives 
$\eta=40\pm8$\%, consistent with our
previous retrograde results.
But what about large-scale magnetic flux? Since its value is
conserved, the differences in the flux cannot be erased by
turbulence; can they affect the outcome?  To test this,  
we carried out a series of simulations 
different only by the magnitude of the magnetic field,
i.e., with $\beta_{\rm min}=\{25, 50, 100,
200\}$.
We refer to these simulations as models A$0.9$N25, A$0.9$N50,
A$0.9$f (our fiducial prograde model), A$0.9$N200, 
respectively. 
The initial fluxes in these models 
differ by a factor of $\approx3$.  Does this lead to a similar difference in
the magnetic flux that reaches the BH?
Figure~\ref{fig:flux} shows
that all models have equatorial magnetic flux profiles, 
$\avg{\phi(r,\theta=\pi/2)}$, that agree to $\lesssim10\%$, 
and $\eta$ that agree to $\lesssim20$\%
(see Table~\ref{tab1}).  
This demonstrates that $\phibh$ and $\eta$
are \emph{independent of the initial
magnetic flux content} of the flow.
We also verified that our results are insensitive to the
initial position of the 
torus  (see model A$0.9$R20 in
Table~\ref{tab1}). 

\begin{figure*}
  \begin{center}
    \includegraphics[height=0.8\columnwidth]{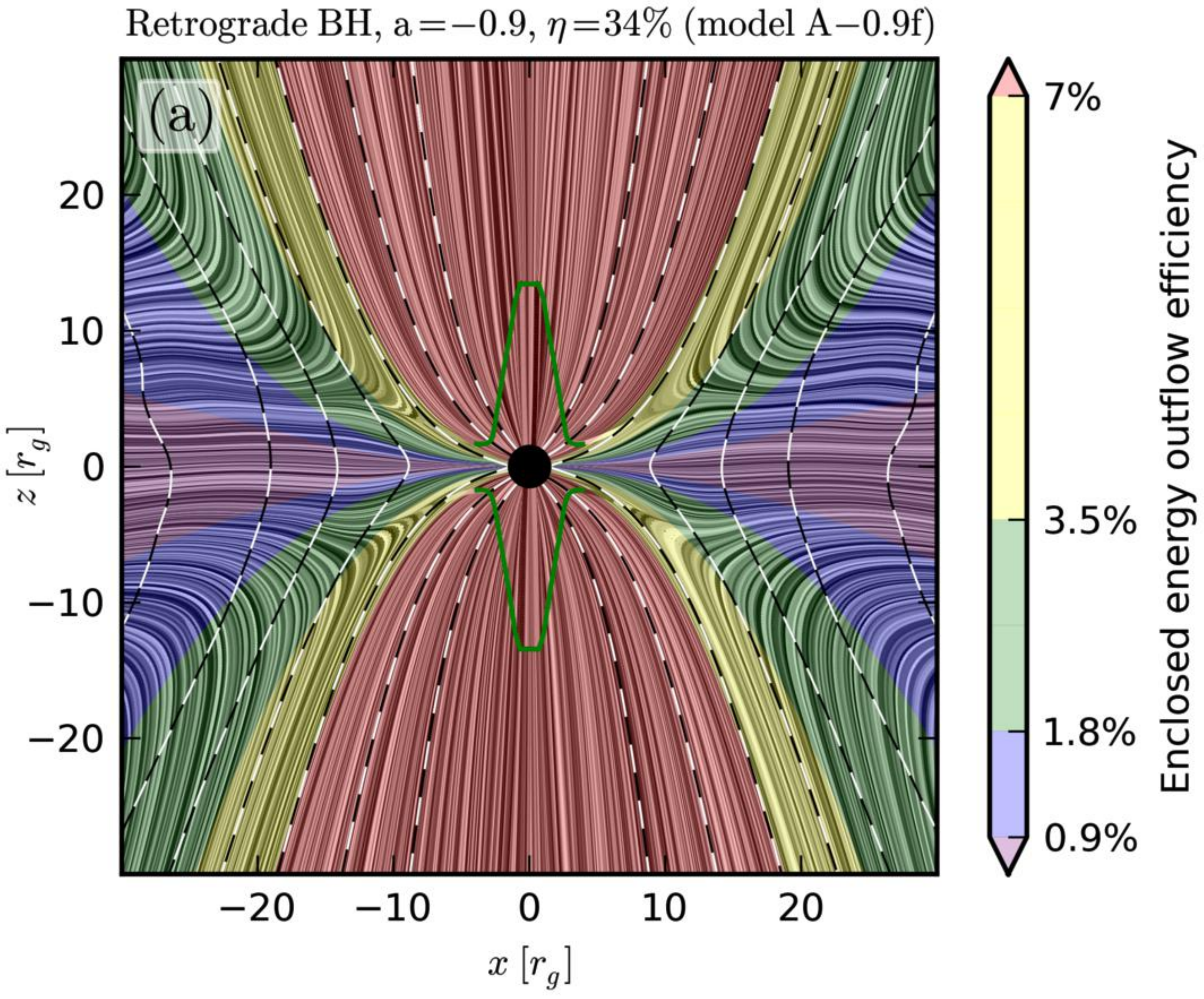}
    \hfill
    \includegraphics[height=0.8\columnwidth]{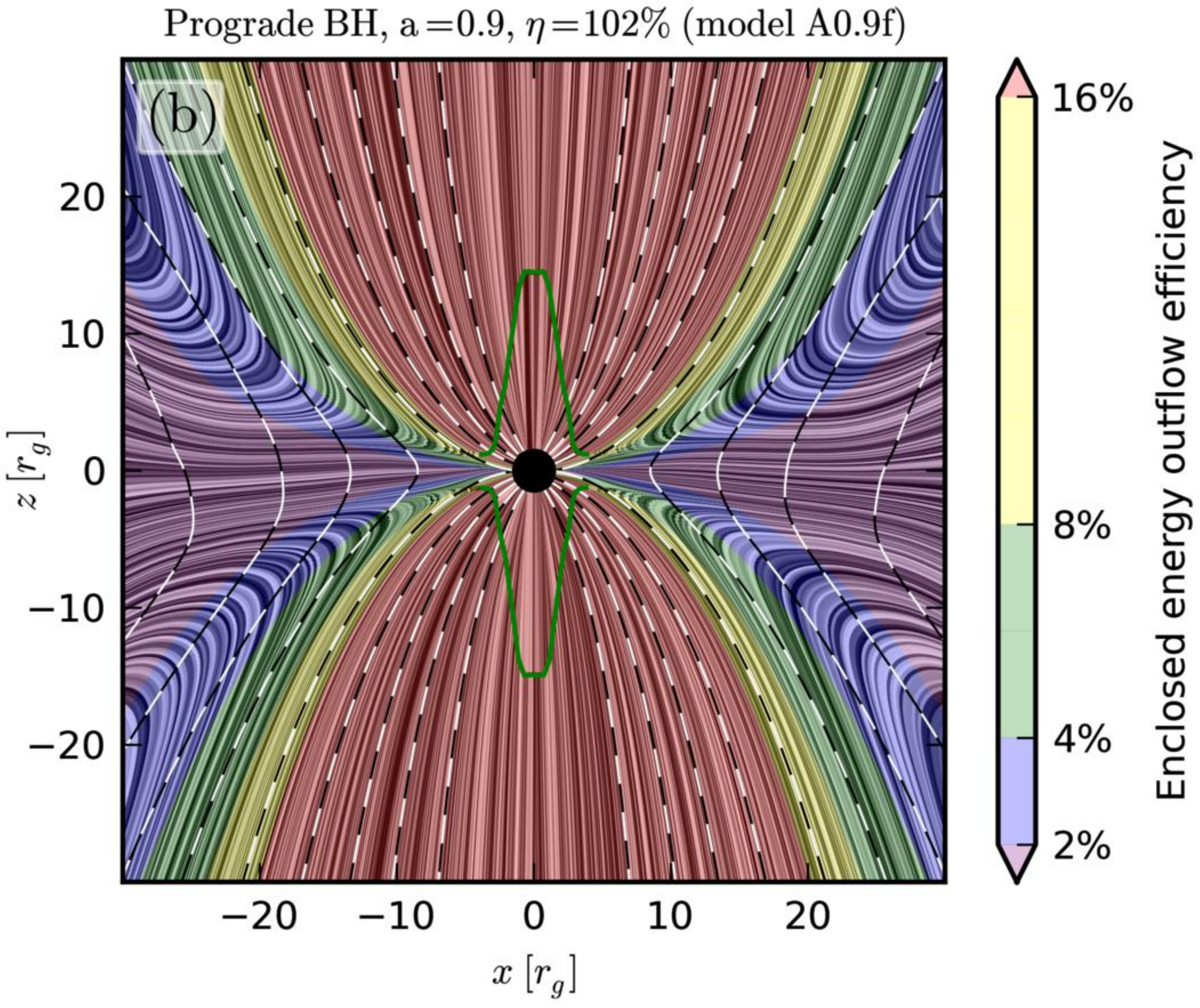}
  \end{center}
  \vspace{-0.2cm}
  \caption{Time and azimuthal average flow pattern in our fiducial
    models, A$-0.9$f (panel a) and A$0.9$f (panel b). BH is
    shown with black filled circle. 
    Black and white dashed lines show poloidal magnetic field lines (constancy levels
    of dimensionless magnetic flux, $\phi=\{10,20,30,...\}$).  Gray
    shading shows velocity streamlines. Green lines
    show the position of stagnation surface, at which $u^r=0$. Colour
    shows energy outflow efficiency enclosed
    between the point in question and an equatorial
    energy flow streamline (see colour bar), so the edges of colour bands
    represent energy flow streamlines.  
    The flow pattern is a standard
    hourglass shape: equatorial inflow at low latitudes, inflow-outflow at
    intermediate latitudes, and twin polar jets (shown in red) at high latitudes. 
}
\label{fig:enflow}
\end{figure*}

So far we considered the total energy output of the BH.  
What fraction of this energy goes into jets and into winds?
The $t$- and $\varphi$-averages
of our fiducial models, A$-0.9$f and A$0.9$f, shown in
Figure~\ref{fig:enflow}, have
similar hourglass shapes. The
equatorial part of the accretion flow directly
reaches the BH, while the inflow at higher
latitudes turns around and forms weakly magnetized 
disc winds.  These winds confine the highly magnetized relativistic
polar jets that directly connect to the BH. 
Figure~\ref{fig:enflow}(a) shows that in the retrograde
model, A$-0.9$f,
the efficiency of the wind is $\etaw(-0.9)\approx7$\%.
The remainder comprises jet efficiency, $\etaj(-0.9)\approx34-7=27\%$. 
Similarly, Figure~\ref{fig:enflow}(b) shows 
that wind efficiency for the prograde model,
A$0.9$f, is $\etaw(0.9)\approx16$\%, so jet efficiency is
$\etaj(0.9)\approx102-16=86$\%. In both cases, the jet, which is
predominantly powered by the BZ mechanism, carries most
($\simeq80$\%) of the total power output, and the power of the wind, 
launched by a BP-like mechanism, is subdominant.
Note that all energy flow streamlines start on the
BH and none start in the body of the disc. This is a manifestation of
energy conservation: energy can either come from the BH via the
release of gravitational binding energy or Penrose/BZ effect, or it can be
advected inward from large distances by accreting material (if the
accretion flow is unbound at large radii).

Our results differ from semi-analytical flux-trapping ``gap'' models
\citep{gar09}, which predict for $\abs{a}=0.9$ that retrograde BHs produce $\gtrsim10\times$
more powerful jets than prograde BHs.  The ``gap'' models make
use of a sharp transition at the ISCO to a plunge flow that sweeps
into the BH all flux from an area (the ``gap'') between the BH horizon
and the ISCO.  Since retrograde BHs have the largest ISCO area, in the
``gap'' model they receive the largest flux and produce the most
powerful jets.  However, poloidal flux distributions in our prograde
and retrograde simulations are strikingly similar and show no ``gap'' (see
Figures \ref{fig:flux}, \ref{fig:enflow}): clearly, the ``gap'' model
does not apply to thick radiatively-inefficient discs (unlike used by \citealt{gar09,ges10}).  This is not surprising
because the sharp transition to the plunge at the ISCO, which the
``gap'' model relies on, is pronounced only for
very thin ($h/r\lesssim0.05$) radiatively-efficient discs (\citealt{penna10}).  In addition, a key piece
of physics is missing from flux-trapping ``gap'' models: ``gap'' models assume that magnetic field is weak and has no back-reaction on
the plunging inflow. However, BH magnetic flux
builds up to a natural saturation point at which the field
does back-react: the magnetic flux not only
modifies the rotation rate of plasma in the plunging region, it also
escapes from the BH via magnetic interchange (\citealt{ss01};
\citetalias{tch11}; \citealt{mtb12}).  
In our simulations, this non-axisymmetric
mechanism fills-in the ISCO region with magnetic flux. No ``gap''
forms because the inflow of magnetic flux happens just as fast as the
outflow of magnetic flux, unlike in the flux-trapping ``gap'' models.
``Gap'' models have not yet accounted for these effects, which
dominate in MADs and, hence, in accretion systems with the most
efficient jets.

\section{Conclusions}
\label{sec:conclusions}

We set out to determine whether prograde or retrograde BHs
produce the most efficient outflows.
Accretion discs with large-scale vertical magnetic fields
naturally 
evolve into a magnetically-arrested
disc (MAD) state in which the central BH is saturated with magnetic flux
and the outflow efficiency ($\eta$) is \emph{maximum}.
Our simulations show that in this state, $\eta$
only depends on the
BH spin, $a$, and the angular thickness of the accretion
disc, $h/r$,
and is \emph{independent of the initial magnetic flux} content of the
disc (see \S\ref{sec:results}).  

We find that 
for medium-thickness discs, $h/r\approx0.3$, and an absolute value of spin,  $\abs{a}=0.9$,
a retrograde BH
produces jets and winds with an efficiency, $\eta=38\pm4$\%, which is
a few times smaller than a
prograde BH, $\eta=106\pm7$\%. 
In both cases, most
of the energy ($\simeq80$\%) emerges
in the form of BZ-powered, relativistic, highly magnetized collimated polar jets (see \S\ref{sec:results}).
We expect the above values of $\eta$ to serve as upper bounds on
jet 
efficiency 
for lower BH spins,
$\abs{a}<0.9$, and thinner discs, $h/r<0.3$.
We also find that a two-fold increase in disc thickness (to
$h/r\approx0.6$) leads to about a three-fold increase in $\eta$
(see \S\ref{sec:results}). 
These results can be used to
place limits on the spin of central BHs from the observed values
of jet efficiency. However,  based on jet
energetics alone it will be challenging to tell apart prograde and retrograde BHs due to their similar efficiencies.
Future studies should consider radiative effects, which are not included
in the current work, and investigate spin-dependence of $\eta$
for thinner discs, with $h/r\lesssim0.1$.

\section*{Acknowledgments}
AT was supported by a Princeton
Center for Theoretical Science Fellowship. AT is grateful to Perimeter
Institute for hospitality.
We thank A.\ Broderick,  D.\ Caprioli, I.\ Contopoulos, D.\ Giannios, L.\ Lehner, M.\
Lyutikov, R.\
Narayan, R.\ Nemmen, D.\ Proga,
J.\ Steiner, J.\ Stone, 
and D.\ Uzdensky for fruitful discussions. We thank the referee, C.\
Reynolds, for useful suggestions.
We acknowledge
NSF support via TeraGrid resources:
NICS Kraken and Nautilus, where simulations were carried out
and data were analyzed, and NCSA MSS and TACC Ranch, where data
were backed up, under grant numbers
TG-AST100040 (AT) 
and TG-AST080025N (JCM).

{\small

\input{ms.bbl}
}






\ifthenelse{\equal{\useiop}{-3}}{
  \bibliographystyle{mn2e} 
}

\section*{Supporting Information}
\label{sec:supportinginfo}
Additional 
Supporting Information may be found in the online version
of this article:
\newline
\hbox{{\bf Movie files.} Movies of models
\href{http://youtu.be/yNZLjsrz0Wo}{A$-0.9$f} and
\href{http://youtu.be/bQE69wti3a4}{A$0.9$f} (click for movies).}

\label{lastpage}
\end{document}

%% file: macros.tex
\newcommand\apj{\rmfamily{ApJ}}%
\newcommand\apjl{\rmfamily{ApJ}}%
\newcommand\apss{\rmfamily{Ap\&SS}}%
\newcommand\aap{\rmfamily{A\&A}}%
\newcommand\mnras{\rmfamily{MNRAS}}%
\newcommand\prd{\rmfamily{Phys.~Rev.~D}}%
\newcommand\pasj{\rmfamily{PASJ}}%
\newcommand\physrep{\rmfamily{Phys.~Rep.}}%

%% file: simtex.txt
         A-0.9f & $-0.9$ &	 $34\pm3$ &	 $100$ &	 $2\pi$ &	 $288\times128\times64$ &	 $15$ &	 $37.1$ &	 $0.7$ &	 $10^5$ &	 $10^3$ &	 $(0; 20060)$ &	 $(8000; 20060)$ \\
          A-0.9 & $-0.9$ &	 $43\pm8$ &	 $100$ &	 $\pi$ &	 $288\times128\times32$ &	 $15$ &	 $37.1$ &	 $0.8$ &	 $10^5$ &	 $10^2$ &	 $(0; 16535)$ &	 $(8000; 16535)$ \\ 
     A-0.9$l_r$ & $-0.9$ &	 $45\pm8$ &	 $100$ &	 $\pi$ &	 $144\times128\times32$ &	 $15$ &	 $37.1$ &	 $0.8$ &	 $10^5$ &	 $10^2$ &	 $(0; 18760)$ &	 $(8000; 18760)$ \\ 
A-0.9$l_\theta$ & $-0.9$ &	 $41\pm4$ &	 $100$ &	 $\pi$ &	 $288\times64\times32$ &	 $15$ &	 $37.1$ &	 $0.8$ &	 $10^5$ &	 $10^2$ &	 $(0; 17570)$ &	 $(8000; 17570)$ \\ 
A-0.9$h_\theta$ & $-0.9$ &	 $40\pm5$ &	 $100$ &	 $\pi$ &	 $288\times256\times32$ &	 $15$ &	 $37.1$ &	 $0.8$ &	 $10^5$ &	 $10^2$ &	 $(12328; 16535)$ &	 $(12328; 16535)$ \\ 
A-0.9$h_\varphi$ & $-0.9$ &	 $36\pm5$ &	 $100$ &	 $\pi$ &	 $288\times128\times64$ &	 $15$ &	 $37.1$ &	 $0.8$ &	 $10^5$ &	 $10^2$ &	 $(8000; 14745)$ &	 $(8000; 14745)$ \\ 
      A-0.9flip & $-0.9$ &	 $40\pm8$ &	 $100$ &	 $\pi$ &	 $288\times128\times32$ &	 $15$ &	 $34.1$ &	 $0.8$ &	 $10^5$ &	 $10^2$ &	 $(14207; 18790)$ &	 $(14207; 18790)$ \\ 
          A0.9f & $0.9$ &	 $102\pm10$ &	 $100$ &	 $2\pi$ &	 $288\times128\times64$ &	 $15$ &	 $34.1$ &	 $0.8$ &	 $10^5$ &	 $10^2$ &	 $(0; 19895)$ &	 $(8000; 19895)$ \\ 
           A0.9 & $0.9$ &	 $96\pm17$ &	 $100$ &	 $\pi$ &	 $288\times128\times32$ &	 $15$ &	 $34.1$ &	 $0.8$ &	 $10^5$ &	 $10^2$ &	 $(0; 22285)$ &	 $(8000; 22285)$ \\ 
      A0.9$l_r$ & $0.9$ &	 $113\pm13$ &	 $100$ &	 $\pi$ &	 $144\times128\times32$ &	 $15$ &	 $34.1$ &	 $0.57$ &	 $10^5$ &	 $10^2$ &	 $(0; 20350)$ &	 $(8000; 20350)$ \\ 
      A0.9$h_r$ & $0.9$ &	 $115\pm20$ &	 $100$ &	 $\pi$ &	 $576\times128\times32$ &	 $15$ &	 $34.1$ &	 $0.8$ &	 $10^5$ &	 $10^2$ &	 $(0; 16265)$ &	 $(8000; 16265)$ \\ 
 A0.9$l_\theta$ & $0.9$ &	 $97\pm14$ &	 $100$ &	 $\pi$ &	 $288\times64\times32$ &	 $15$ &	 $34.1$ &	 $0.8$ &	 $10^5$ &	 $10^2$ &	 $(0; 16915)$ &	 $(8000; 16915)$ \\ 
 A0.9$h_\theta$ & $0.9$ &	 $123\pm22$ &	 $100$ &	 $\pi$ &	 $288\times256\times32$ &	 $15$ &	 $34.1$ &	 $0.8$ &	 $10^5$ &	 $10^2$ &	 $(14207; 22290)$ &	 $(14207; 22290)$ \\ 
A0.9$h_{\theta}h_{\varphi}$ & $0.9$ &	 $112\pm19$ &	 $100$ &	 $\pi$ &	 $288\times256\times64$ &	 $15$ &	 $34.1$ &	 $0.8$ &	 $10^5$ &	 $10^2$ &	 $(14207; 24410)$ &	 $(14207; 24410)$ \\ 
A0.9$h_\varphi$ & $0.9$ &	 $109\pm21$ &	 $100$ &	 $\pi$ &	 $288\times128\times64$ &	 $15$ &	 $34.1$ &	 $0.8$ &	 $10^5$ &	 $10^2$ &	 $(0; 15550)$ &	 $(8000; 15550)$ \\ 
A0.9$h^2_\varphi$ & $0.9$ &	 $103\pm14$ &	 $100$ &	 $\pi$ &	 $288\times128\times128$ &	 $15$ &	 $34.1$ &	 $0.8$ &	 $10^5$ &	 $10^2$ &	 $(8500; 14625)$ &	 $(8500; 14625)$ \\ 
        A0.9N25 & $0.9$ &	 $116\pm32$ &	 $25$ &	 $\pi$ &	 $288\times128\times32$ &	 $15$ &	 $34.1$ &	 $0.8$ &	 $10^5$ &	 $10^2$ &	 $(0; 17350)$ &	 $(8000; 17350)$ \\ 
        A0.9N50 & $0.9$ &	 $114\pm20$ &	 $50$ &	 $\pi$ &	 $288\times128\times32$ &	 $15$ &	 $34.1$ &	 $0.8$ &	 $10^5$ &	 $10^2$ &	 $(0; 14385)$ &	 $(8000; 14385)$ \\ 
       A0.9N200 & $0.9$ &	 $118\pm32$ &	 $200$ &	 $\pi$ &	 $288\times128\times32$ &	 $15$ &	 $34.1$ &	 $0.8$ &	 $10^5$ &	 $10^2$ &	 $(0; 28620)$ &	 $(16000; 28620)$ \\ 
        A0.9R20 & $0.9$ &	 $102\pm19$ &	 $100$ &	 $\pi$ &	 $288\times128\times32$ &	 $20$ &	 $45.35$ &	 $0.8$ &	 $10^5$ &	 $10^2$ &	 $(0; 15295)$ &	 $(8000; 15295)$ \\ 

%% file: simtex_thick.txt
    A-0.94BfN30 & $-0.9375$ &	 $88\pm75$ &	 $30$ &	 $2\pi$ &	 $136\times64\times128$ &	 $10$ &	 $100$ &	 $0.73$ &	 $26000$ &	 $500$ &	 $(0; 13000)$ &	 $(8000; 13000)$ \\ 
     A0.94BfN30 & $0.9375$ &	 $354\pm43$ &	 $30$ &	 $2\pi$ &	 $136\times64\times128$ &	 $10$ &	 $100$ &	 $0.73$ &	 $26000$ &	 $500$ &	 $(0; 13000)$ &	 $(8000; 13000)$ \\ 